\documentclass[11pt,letterpaper]{article}
\usepackage[T1]{fontenc}
\usepackage[utf8]{inputenc}
\usepackage[margin=1in]{geometry}
\usepackage{newtxtext}
\usepackage{amsmath,amsthm}
\usepackage{newtxmath}
\usepackage{bm,bbm}
\usepackage{graphicx,booktabs,multirow}
\usepackage{microtype}
\usepackage{enumitem}
\usepackage[authoryear,round]{natbib}
\usepackage[font=small,labelfont=bf]{caption}
\usepackage{xcolor}
\usepackage[colorlinks=true,allcolors=blue]{hyperref}
\usepackage{cleveref}
\hypersetup{
  pdftitle={Beyond Exchangeability: Distribution-Shift-Aware Integration of External Control Data in Randomized Trials},
  pdfauthor={Jiawei Shan, Yiteng Tu, Guanbo Wang, Chao Ying, Jiwei Zhao}
}
\AtBeginDocument{}
\allowdisplaybreaks
\newtheorem{assumption}{Assumption}

\crefname{assumption}{Assumption}{Assumptions}
\Crefname{assumption}{Assumption}{Assumptions}
\newtheorem{Lem}{Lemma}

\crefname{Lem}{Lemma}{Lemmas}
\Crefname{Lem}{Lemma}{Lemmas}
\newtheorem{Pro}{Proposition}

\crefname{Pro}{Proposition}{Propositions}
\Crefname{Pro}{Proposition}{Propositions}
\newtheorem{Th}{Theorem}

\crefname{Th}{Theorem}{Theorems}
\Crefname{Th}{Theorem}{Theorems}
\newtheorem{Cor}{Corollary}

\crefname{Cor}{Corollary}{Corollaries}
\Crefname{Cor}{Corollary}{Corollaries}
\theoremstyle{definition}
\newtheorem{Rem}{Remark}

\crefname{Rem}{Remark}{Remarks}
\Crefname{Rem}{Remark}{Remarks}

\def\X{{\bf X}}
\def\x{{\bf x}}
\def\E{{\mathrm{E}}}
\def\O{{\mathcal{O}}}
\def\calX{{\cal X}}
\def\calY{{\cal Y}}
\def\calR{{\cal R}}
\def\calN{{\cal N}}
\def\mR{\mathbb{R}}
\def\calV{{\mathcal V}}
\def\ba{{\boldsymbol\alpha}}
\def\bc{{\boldsymbol c}}
\def\bb{{\boldsymbol\beta}}
\def\bg{\boldsymbol\gamma}
\def\bphi{\boldsymbol\phi}
\def\bSigma{{\boldsymbol\Sigma}}
\def\0{{\bf 0}}
\def\trans{^{\rm T}}
\def\wh{\widehat}
\def\wt{\widetilde}
\def\var{\hbox{var}}
\def\cov{\hbox{cov}}
\def\argmin{\mbox{argmin}}
\newcommand{\convd}{\xrightarrow{d}}
\newcommand{\convp}{\xrightarrow{p}}
\def\gs{{\rm gs}}
\DeclareMathOperator{\pr}{pr}
\newcommand{\ind}{\mathrel{\perp\!\!\!\perp}}

\begin{document}

\title{Beyond Exchangeability: Distribution-Shift-Aware Integration of External Control Data in Randomized Trials}
\author{
Jiawei Shan$^{1}$ \quad Yiteng Tu$^{2}$ \quad Guanbo Wang$^{3}$ \quad Chao Ying$^{1}$ \quad Jiwei Zhao$^{1}$\thanks{E-mail: \href{mailto:jiwei.zhao@wisc.edu}{\texttt{jiwei.zhao@wisc.edu}}.}\\[0.7em]
\small $^{1}$University of Wisconsin-Madison\\
\small $^{2}$University of California, San Diego\\
\small $^{3}$Dartmouth College
}
\date{}
\maketitle

\begin{abstract}
Randomized controlled trials (RCTs) are the gold standard for evaluating causal effects but are often costly and difficult to scale; consequently, they are frequently augmented with auxiliary external controls in many applications. Prior approaches to borrowing such data typically rely on exchangeability, under which the external controls are readily usable for inference in the trial population. In practice, however, differences in eligibility criteria, standard of care, and data collection procedures may induce distribution shifts between the RCT and the external controls, making exchangeability implausible. In this paper, we propose a general framework for incorporating external controls that explicitly accounts for distribution shifts through an augmented estimating function, allowing information about the shifts to be leveraged for efficiency gains even when exchangeability does not hold.
To guard against potential bias and efficiency loss due to misspecification of the distribution shift, we further develop an adaptive shrinkage estimator that preserves consistency while guaranteeing efficiency dominance over the trial-only benchmark. Synthetic experiments and a real data application demonstrate the practical advantages of the proposed approaches.
\end{abstract}

\section{Introduction}

Randomized controlled trials (RCTs) are the gold standard for evaluating treatment effects because randomization removes confounding bias by design \citep{ImbensRubin2015}. In practice, however, RCTs often produce estimates that are not sufficiently precise, due to optimistic assumptions about event rates, follow-up, and accrual, as well as limited power for subgroup analyses. This has motivated growing interest in augmenting RCTs with auxiliary external data, such as historical trials, hospital registries, electronic health records, and claims \citep{Pocock1976,ColnetMayerChenDiengEtAl2024}. In many studies of experimental interventions, these auxiliary data contain only untreated subjects and are therefore used as external controls (ECs). A straightforward way to incorporate ECs is to assume \emph{exchangeability}: conditional on observed covariates, the untreated potential outcome distribution, or some more essential characteristics such as its conditional mean, is comparable between the trial and EC populations. Under this assumption, ECs are readily usable for trial inference, and many methods have been developed to borrow such external information \citep{DahabrehRobertsonTchetgenStuartEtAl2019,LiMiaoLuZhou2023,ValanciusPangZhuColeEtAl2024}.

In practice, however, exchangeability is often tenuous. Differences in eligibility criteria, standards of care, calendar time, recruitment mechanisms, and measurement protocols can induce distribution shifts that make ECs systematically different from trial controls. Recent work has therefore developed guarded approaches aiming to reduce the risk of biased borrowing by selecting compatible external subjects or by downweighting, discounting, or shrinking external information \citep{RosenmanBasseOwenBaiocchi2023,YangGaoZengWang2023,GaoYangShanYEEtAl2024}. These strategies improve robustness but still rely on the compatibility of the available EC data, and may discard or attenuate useful information. Moreover, existing methods typically borrow EC information only for the control arm, overlooking the possibility that external covariate information, when properly calibrated to the trial population, can improve inference for both the treatment and control arms.

To tackle these challenges, we develop a general framework for integrating ECs into trial inference that explicitly accounts for distribution shifts, thereby allowing available information about discrepancies between the trial and EC populations to be incorporated for potential efficiency gains without requiring exchangeability. Specifically, we consider both \emph{covariate} shift, which captures differences in covariate distributions, and \emph{concept} shift, which captures differences in conditional outcome distributions. We derive calibration equations that explicitly involve these shifts and use them to augment trial-only efficient influence functions, yielding EC-augmented estimating equations for both the treatment and control arms that can draw on shift information from sources such as design knowledge, previously linked studies, or validation data.

Importantly, EC data alone do not necessarily improve efficiency. When no additional information about the distribution shifts is available, estimating unrestricted shift models solely from the analysis data need not yield efficiency gains. Moreover, misspecification of the shift models and variability in their estimation may introduce bias or render the augmented estimator less efficient than trial-only inference. To guard against these risks, we further develop a data-adaptive shrinkage estimator that combines the EC-augmented estimator with a trial-only benchmark. The shrinkage estimator preserves trial-based consistency and guarantees asymptotic efficiency dominance over the trial-only benchmark, while retaining the potential efficiency gains when the available shift information is sufficiently informative and reliable.

To summarize, we make the following novel contributions in this work:
\begin{itemize}[leftmargin=*]
    \item We introduce a new framework for integrating EC data into RCT analysis without assuming exchangeability, by explicitly exploiting distribution shifts between the trial and EC populations.
    \item We propose calibration-based augmented estimating functions from trial-only efficient influence functions, exploiting both covariate and concept shifts so that EC data can improve inference for both treated and control arms.
    \item We further propose a data-adaptive shrinkage estimator that remains consistent under shift-model misspecification while guaranteeing efficiency dominance over the trial-only benchmark.
\end{itemize}
We have conducted extensive synthetic experiments and a real-data application to demonstrate the finite-sample advantages of the proposed approaches.

\section{Preliminary}

Let $\X\in\calX\subset\mR^p$ denote the pre-treatment covariates and $Y\in\calY\subset\mR$ the outcome of interest.
Let $T\in\{0,1\}$ denote the treatment assignment, with $T=1$ for treatment and $T=0$ for control. We adopt the potential outcome framework $Y=T Y_1 + (1-T) Y_0$, where $Y_1$ and $Y_0$ denote the potential outcomes under treatment and control, respectively.

Suppose we have access to data from an RCT and an external control cohort that contains only untreated subjects. Let $R$ indicate the data source, where $R=1$ for RCT and $R=0$ for EC. The observed data $\O$ consist of $N$ independent and identically distributed (i.i.d.) samples of $(R,T,\X,Y)$ from the combined population, partitioned into the RCT sample, $\O_1=\{(r_i=1,t_i,\x_i,y_i),\, i=1,\dots,n\}$, and the EC sample, $\O_2=\{(r_i=0,t_i=0,\x_i,y_i),\, i=n+1,\dots,N\}$. Assume that $n/N \to \kappa\in(0,1)$ as $n,N\to\infty$.

The causal estimand of interest is the average treatment effect (ATE) in the trial population, defined as $\tau=\tau_1-\tau_0$, where $\tau_t=\E(Y_t\mid R=1)$ for $t\in\{0,1\}$. The following standard assumptions, which formalize the RCT design, ensure identification of the causal parameters \citep{RosenbaumRubin1983}:
\begin{assumption}\label{ass:rct_ec}
    (i) $T\ind (\X,Y_0,Y_1)\mid R=1$;
    (ii) $0<\pi=\pr(T=1\mid R=1)<1$.
\end{assumption}

Under \Cref{ass:rct_ec}, $\tau_1$ and $\tau_0$ are identified from the trial data $\O_1$. A natural pair of gold-standard estimators is given by the treated and control sample means $\wt\tau_1^\gs = \sum_{i=1}^{n}t_i y_i/\sum_{i=1}^{n}t_i$ and $\wt\tau_0^\gs = \sum_{i=1}^{n}(1-t_i) y_i/\sum_{i=1}^{n}(1-t_i)$.
To characterize the most efficient estimator based on the trial data, define $\mu_t(\x)=\E(Y\mid \x,t,R=1)$ for $t\in\{0,1\}$.
The efficient influence functions (EIF) \citep{RobinsRotnitzkyZhao1994} for estimating $\tau_1$ and $\tau_0$, are, respectively,
\begin{align}
&\wt\psi_1 = \frac{RT}{\kappa\pi}\{Y-\mu_1(\X)\}+ \frac{R}{\kappa}\left\{\mu_1(\X)-\tau_1\right\},\label{eq:EIF1_rct}\\
\mbox{and}~~
&\wt\psi_0 = \frac{R(1-T)}{\kappa(1-\pi)}\{Y-\mu_0(\X)\}+ \frac{R}{\kappa}\left\{\mu_0(\X)-\tau_0\right\}.\label{eq:EIF0_rct}
\end{align}
Denote $\wt\psi=\wt\psi_1-\wt\psi_0$. Thus, the semiparametric efficiency bound for estimating $\tau$ based on the trial data solely is $\wt B=\E(\wt\psi^2)$.

\section{Exploitation of Distribution Shifts between RCT and EC}\label{sec:exploitation}

A straightforward way to integrate EC data in RCT is to assume certain similarity between the two sources, which was termed \emph{exchangeability} in the literature.
For instance, one may assume the distribution exchangeability, namely,
\begin{align*}
p(y_0 \mid \x)=q(y_0 \mid \x),
\end{align*}
where $p(\cdot)$ and $q(\cdot)$ denote the distribution in the RCT and the EC populations, respectively, throughout the paper.
This certainly implies that the conditional mean of $Y_0$ is shared across the trial and the external controls, namely,
$\E(Y_0 \mid \x, R=1)=\E(Y_0 \mid \x, R=0)$, i.e., the mean exchangeablity.
Accordingly, the external controls can be readily used to inform untreated outcomes in the trial population.

However, this type of strong similarity assumption may often fail in applications because it is the rule rather than the exception that the distributions between RCT and EC differ.
To move beyond exchangeability, one needs to explicitly model the discrepancy between the two populations.
Let $a(\x):=q(\x)/p(\x)$ represent the \emph{covariate} shift between the two data sources.
Further, let $b(\x,y_0):=q(y_0\mid\x)/p(y_0\mid\x)$ represent the \emph{concept} shift in the sense of the conditional distribution of the untreated potential outcome.
To ease the presentation, we define
\begin{equation}\label{eq:def_k_rho}
  \begin{aligned}
  k(\x)&:=\pr(R=1\mid\x) = \frac{\kappa}{\kappa+(1-\kappa)a(\x)},\\
  \mbox{and}~~
   \rho(\x,y_0) &:= \pr(R=1\mid \x,y_0,T=0) = \frac{\kappa(1-\pi)}{\kappa(1-\pi)+(1-\kappa)a(\x)b(\x,y_0)}.
  \end{aligned}
\end{equation}
The following lemma shows how $k(\x)$ and $\rho(\x,y_0)$ can be used to balance distributions across the trial, external controls, and the combined population.
\begin{Lem}\label{lemma:tranfer_new}
 Under \Cref{ass:rct_ec}, for any integrable functions $g_1(\x)$ and $g_2(\x,y_0)$, it holds that
  \begin{align*}
  \E\left\{g_1(\X)\mid R=1\right\}
    &= \E\left\{\frac{k(\X)}{\kappa}g_1(\X)\right\}, \mbox{ and }\\
  \E\left\{g_2(\X,Y_0)\mid R=1\right\}
  &=   \E\left\{\frac{1-T}{\kappa(1-\pi)}\rho(\X,Y_0)g_2(\X,Y_0)\right\} .
  \end{align*}
\end{Lem}
\Cref{lemma:tranfer_new} motivates the idea of augmenting the trial-only estimating functions, $\wt\psi_1$ and $\wt\psi_0$, to incorporate external controls by exploiting and modeling the distributional shifts. For $\tau_1$, the EC sample contributes only covariate information through $\X$, yielding the following EC-augmented estimating function:
   \begin{align}\label{eq:IF1_prop}
        \psi_1 =
        \frac{RT}{\kappa\pi}\{Y-\mu_1(\X)\}
        + \frac{k(\X)}{\kappa}\{\mu_1(\X)-\tau_1\}.
    \end{align}
For $\tau_0$, the EC sample provides information on both covariates and untreated outcomes, leading to the following augmented estimating function:
\begin{align}\label{eq:IF0_prop}
      \psi_0 =
       \frac{1-T}{\kappa(1-\pi)} \rho(\X,Y) \{Y-\mu_0(\X)\}
      + \frac{k(\X)}{\kappa}\{\mu_0(\X)-\tau_0\}.
  \end{align}
Let $\Delta(\x)=\mu_1(\x)-\mu_0(\x)$ and let $\psi=\psi_1-\psi_0$ be the estimating function of $\tau$.
The following proposition establishes the validity of $\psi_t$ and characterizes its potential efficiency improvement over the trial-only estimating function $\wt\psi_t$.
\begin{Pro}\label{prop:var}
   Under \Cref{ass:rct_ec}, we have $\E(\psi_t)=\E(\psi)=0$ for $t\in\{0,1\}$. Let
  \begin{align*}
    \calV_\rho = \frac{1}{\kappa^2(1-\pi)^2}\E\left[(1-T)\{Y-\mu_0(\X)\}^2 \rho(\X,Y)\left\{1-\rho(\X,Y)\right\}\right] \ge 0,
  \end{align*}
  denote the contribution of the concept shift. Then
  \begin{align*}
    \E(\wt\psi_1^2) - \E(\psi_1^2) &=  \frac{1}{\kappa^2}\E\left[\{\mu_1(\X)-\tau_1\}^2 k(\X)\left\{1-k(\X)\right\}\right]\ge0,\\
    \E(\wt\psi_0^2) - \E(\psi_0^2) &=  \frac{1}{\kappa^2}\E\left[\{\mu_0(\X)-\tau_0\}^2 k(\X)\left\{1-k(\X)\right\}\right] + \calV_\rho\ge0,\\
    \mbox{and}\quad
    \E(\wt\psi^2) - \E(\psi^2) &=  \frac{1}{\kappa^2}\E\left[\{\Delta(\X)-\tau\}^2 k(\X)\left\{1-k(\X)\right\}\right] + \calV_\rho\ge0.
  \end{align*}
    Each difference is strictly positive as long as there exists a set of $\x$ with nonzero measure on which the densities overlap, i.e., $p(\x)q(\x)>0$, and the relevant mean function, $\mu_t(\x)$ or $\Delta(\x)$, is non-constant. An analogous condition applies to $\calV_\rho$.
\end{Pro}

\section{Proposed Augmented Estimator Incorporating EC Data}\label{sec:augmented}

The augmented estimating functions \eqref{eq:IF1_prop}-\eqref{eq:IF0_prop} explicitly account for distribution shifts through the weights $k$ and $\rho$, thereby incorporating external controls without assuming exchangeability. In this section, we focus on the corresponding augmented estimators under parametric working models $k(\x;\ba)$, $\rho(\x,y_0;\bb)$, and $\mu_t(\x;\bg_t)$ for $k(\x)$, $\rho(\x,y_0)$, and $\mu_t(\x)$, respectively, and study their asymptotic properties. This also provides a concrete illustration of how estimating the distribution shifts affects the resulting treatment effect estimators.

Let $(\wh\ba,\wh\bb,\wh\bg_0,\wh\bg_1)$ denote $N^{1/2}$-consistent estimators of the nuisance parameters, with a probability limit $(\ba^*,\bb^*,\bg^*_0,\bg^*_1)$.
Then, an estimator of $\tau_1$ that solves \eqref{eq:IF1_prop} is
\begin{align*}
\wh\tau_1 =  \frac{1}{N}\sum_{i=1}^{N} \left[\frac{r_tt_i}{\wh\pi}\{y_i-\mu_1(\x_i;\wh\bg_1)\}
+  k(\x_i;\wh\ba)\mu_1(\x_i;\wh\bg_1)\right]\bigg/ \frac{1}{N}\sum_{i=1}^{N}k(\x_i;\wh\ba),
\end{align*}
and an estimator of $\tau_0$ that solves \eqref{eq:IF0_prop} is
\begin{align*}
\wh\tau_0 =  \frac{1}{N}\sum_{i=1}^{N} \left[\frac{1-t_i}{1-\wh\pi} \rho(y_i,\x_i;\wh\bb) \{y_i-\mu_0(\x_i;\wh\bg_0)\}
+  k(\x_i;\wh\ba)\mu_0(\x_i;\wh\bg_0)\right]\bigg/ \frac{1}{N}\sum_{i=1}^{N}k(\x_i;\wh\ba),
\end{align*}
where $\wh\pi=\#\{T=1\}/n$. An estimator of the treatment effect $\tau$ is then given by $\wh\tau=\wh\tau_1-\wh\tau_0$.

Compared with the trial-based estimators solving \eqref{eq:EIF1_rct}--\eqref{eq:EIF0_rct}, the augmented estimator $\wh\tau_1$ entails an extra working model $k(\x; \ba)$, while $\wh\tau_0$ entails extra working models $k(\x; \ba)$ and $\rho(\x,y_0; \bb)$ to incorporate external controls. The following \Cref{thm:consistency}
establishes the consistency and asymptotic normality of the augmented estimators.

\begin{Th}\label{thm:consistency}
  Under \Cref{ass:rct_ec} and regularity conditions in Theorems 2.6 and 3.4 of \cite{NeweyMcFadden1994}, the estimator $\wh\tau_1$ is consistent and asymptotically normal (CAN) if $k(\x; \ba^*)=k(\x)$. Moreover, the estimator $\wh\tau_0$ is CAN if $k(\x; \ba^*)=k(\x)$ and $\rho(y_0,\x; \bb^*)=\rho(y_0,\x)$.
\end{Th}

\Cref{thm:consistency} shows that consistency of $\wh\tau_1$ relies on correct model specification of $k(\x)$, while remaining robust to potential misspecification of $\mu_1(\x)$. Likewise, the consistency of $\wh\tau_0$ relies on correct model specification of $k(\x)$ and $\rho(y_0,\x)$, while remaining robust to potential misspecification of $\mu_0(\x)$. As shown in \eqref{eq:def_k_rho}, correct specification of $k(\x)$ and $\rho(y_0,\x)$ is equivalent to a ``precise'' characterization of the distribution shifts $a(\x)$ and $b(\x,y_0)$. Because the augmented estimating functions incorporate external data through the modeled distributional shifts, consistency naturally hinges on correct specification of those shifts. At the same time, the augmented estimator $\wh\tau_t$ retains robustness to possible misspecification of the outcome models $\mu_t(\x)$, mirroring the robustness property enjoyed by trial-only doubly robust estimators from \eqref{eq:EIF1_rct}-\eqref{eq:EIF0_rct}.
The following \Cref{thm:asy_dist} further presents the asymptotic representation when all working models are correctly specified.

\begin{Th}\label{thm:asy_dist}
  Suppose $k(\x; \ba^*)=k(\x)$, $\rho(y_0,\x; \bb^*)=\rho(y_0,\x)$, $\mu_t(\x; \bg_t^*)=\mu_t(\x)$ for $t\in\{0,1\}$, and assumptions in \Cref{thm:consistency} hold. Define
  \begin{align*}
    \Gamma_t =  \E\left[\frac{\{\mu_t(\X)-\tau_t\}}{\kappa}\frac{\partial k(\X;\ba^*)}{\partial{\ba}\trans}\right],
    \mbox{ and }
    \Gamma_\rho =  \E\left[\frac{1-T}{\kappa(1-\pi)}\{Y_0-\mu_0(\X)\}\frac{\partial\rho(\X,Y_0;\bb^*)}{\partial{\bb}\trans}\right],
  \end{align*}
  for $t\in\{0,1\}$. Then, we have \\
  (i) $\sqrt{N}(\wh\tau_1-\tau_1)=N^{-1/2}\sum_{i=1}^{N}\psi_{1,i}+\calR_1+o_p(1)$, where $\calR_1=\Gamma_1\cdot \sqrt{N}(\wh\ba-\ba^*)$;\\
  (ii) $\sqrt{N}(\wh\tau_0-\tau_0)=N^{-1/2}\sum_{i=1}^{N}\psi_{0,i}+\calR_2+\calR_3+o_p(1)$, where $\calR_2=\Gamma_0\cdot \sqrt{N}(\wh\ba-\ba^*)$ and $\calR_3=\Gamma_\rho\cdot \sqrt{N}(\wh\bb-\bb^*)$.
\end{Th}

\Cref{thm:asy_dist} shows that, under correct model specification, estimating $(\bg_1,\bg_0)$ does not jeopardize the asymptotic behavior of the treatment effect estimators $(\wh\tau_1,\wh\tau_0)$, though estimating $(\ba,\bb)$ does.
Indeed, this well aligns with the intuition that some prior knowledge about how distribution shifts helps gaining the efficiency.
The more one knows, the more efficient the estimators are.
An ideal scenario arises when the distribution shifts are completely known from the study design or reliable external knowledge, so that the weights $k$ and $\rho$ need not be estimated. In this case, the remainder terms $\calR_1$, $\calR_2$, and $\calR_3$ in \Cref{thm:asy_dist} vanish, yielding
    $\sqrt{N}(\wh\tau_1-\tau_1)
     \convd \calN\{0,\E(\psi_1^2)\}$,
     and
    $\sqrt{N}(\wh\tau_0-\tau_0)
     \convd \calN\{0,\E(\psi_0^2)\}$.
  Thus, by \Cref{prop:var}, the augmented estimators guarantee an efficiency gain over trial-only inference.
  When the shifts are only partially known, so that the working models for $k$ and $\rho$ contain fewer free parameters, efficiency gains are also plausible. For example, \citet{KimYu2011} consider an exponential tilting model for outcome-dependent selection, $q(y_0\mid\x)\propto \exp(\theta y_0)\,p(y_0\mid\x)$, with $\theta$ known a priori; see \Cref{sec:simulation} for a numerical illustration.

In practice, it is also of interest to consider a validation setting in which an auxiliary validation dataset is available for estimating the shifts.
Suppose an independent validation sample of size $m$ is available for estimating $(\ba,\bb)$, with $m/N\to \xi\in(0,\infty)$, and $ \sqrt{m} \{(\wh\ba-\ba^*)\trans,(\wh\bb-\bb^*)\trans\}\trans= m^{-1/2}\sum_{j=1}^{m}\bphi_j+o_p(1)$,
  where $\bphi_j=(\bphi_{\alpha,j}\trans,\bphi_{\beta,j}\trans)\trans$ and $\E(\bphi)=\0$. Then, \Cref{thm:asy_dist} implies that $\sqrt{N}(\wh\tau_1-\tau_1)\convd \calN(0,\sigma_{1,\rm val}^2)$ and $\sqrt{N}(\wh\tau_0-\tau_0)\convd \calN(0,\sigma_{0,\rm val}^2)$, where
  \begin{align*}
    \sigma_{1,\rm val}^2 =  \E(\psi_1^2)+ \xi^{-1}\Gamma_1\var(\bphi_\alpha)\Gamma_1\trans,
    \mbox{ and }
    \sigma_{0,\rm val}^2 =  \E(\psi_0^2)+ \xi^{-1}(\Gamma_0,\Gamma_\rho)\var(\bphi)(\Gamma_0,\Gamma_\rho)\trans.
  \end{align*}
  Therefore, the additional variance induced by estimating the distribution shift from validation data is of order $\xi^{-1}$. As the validation sample becomes large relative to the primary sample, the extra terms vanish, and the asymptotic variances converge to $\E(\psi_1^2)$ and $\E(\psi_0^2)$, which are exactly the variances obtained when the distribution shift is completely known.

\section{Proposed Shrinkage Estimator with Guaranteed Consistency and Efficiency Dominance}\label{sec:shrinkage}

In applications, the EC-augmented estimators proposed in Section~\ref{sec:augmented} may suffer from inconsistency or loss of efficiency because of the misspecification of the shift model or the additional variability from sufficiently estimating it. To address this issue, we propose a dependable, data-adaptive shrinkage estimator that fuses $\wh\tau_t$ (potentially more efficient but misspecification-sensitive) with the trial-only estimator $\wt\tau_t$ (guaranteed consistency but potentially less efficient), to guarantee consistency while also retaining potential efficiency gains, even under model misspecification.

We consider estimators of the form
\begin{align*}
\wh\tau_t(\lambda_t) = \wt\tau_t + \lambda_t (\wh\tau_t - \wt\tau_t),~~\lambda_t\in\mR,
\end{align*}
where $\lambda_t$ is the shrinkage weight. Note that $\wh\tau_t(0)=\wt\tau_t$ and $\wh\tau_t(1)=\wh\tau_t$.
The key insight is as follows. If $\wh\tau_t$ is inconsistent (i.e., $\wh\tau_t\not\to\tau_t$), we expect $\lambda_t$ to shrink toward zero, so that the combined estimator reduces to the consistent baseline $\wt\tau_t$. When $\wh\tau_t$ is consistent for $\tau_t$, $\wh\tau_t(\lambda_t)$ retains consistency for any choice of $\lambda_t$. In this case, we expect $\lambda_t$ to converge to an optimal $\lambda_t^*$ that minimizes the asymptotic variance of $\wh\tau_t(\lambda_t)$, ensuring that it is at least as efficient as either separate estimator.

We first calculate $\lambda_t^*$.
Noting that $\var\{\wh\tau_t(\lambda_t)\} = \var(\wt\tau_t)+2\lambda_t\cov(\wh\tau_t - \wt\tau_t,\wt\tau_t)+\lambda_t^2 \var(\wh\tau_t - \wt\tau_t)$, we have
\begin{align*}
\lambda_t^*=\argmin_{\lambda_t} \var\{\wh\tau_t(\lambda_t)\} = -\frac{\cov(\wh\tau_t - \wt\tau_t,\wt\tau_t)}{\var(\wh\tau_t - \wt\tau_t)}.
\end{align*}
It is straightforward to verify that
\begin{align*}
\sigma^{2}_t := \var\{\wh\tau_t(\lambda_t^*)\} = \frac{\var(\wt\tau_t)\var(\wh\tau_t)-\cov(\wh\tau_t,\wt\tau_t)^2}{\var(\wh\tau_t - \wt\tau_t)} \le \min \{\var(\wt\tau_t),\var(\wh\tau_t)\}.
\end{align*}
To facilitate adaptive shrinkage driven by the consistency of $\wh\tau_t$, we define, with $V_{t,N}:=\var(\wh\tau_t - \wt\tau_t)$, and a given $\gamma\in(0,3/2)$,
\begin{align*}
\delta_{t,N} = \frac{V_{t,N}^2}{V_{t,N}^2+(\wh\tau_t - \wt\tau_t)^4/N^\gamma}
= \left\{1+N^{-\gamma}\left(\frac{\wh\tau_t - \wt\tau_t}{\sqrt{V_{t,N}}}\right)^{4}\right\}^{-1}.
\end{align*}
Noting that $V_{t,N}=O(N^{-1})$, if $\wh\tau_t$ is consistent, then $\wh\tau_t - \wt\tau_t = O_p(N^{-1/2})$ and hence $\delta_{t,N}=1+O_p(N^{-\gamma})$. Otherwise, if $\wh\tau_t$ is inconsistent, i.e., $\wh\tau_t - \wt\tau_t = c^*_t + O_p(N^{-1/2})$ for some $c^*_t\neq 0$, we have $\delta_{t,N}=O_p(N^{\gamma-2})=o_p(N^{-1/2})$. Denote $\lambda_{t,N}=\delta_{t,N}\lambda_t^*$, then we have
\begin{align*}
  \lambda_{t,N} = \left\{\begin{aligned}
    & \lambda_t^* + O_p(N^{-\gamma}),~~\mbox{if}~\wh\tau_t\convp\tau_t, \\
    & o_p(N^{-1/2}),~~\mbox{otherwise}.\\
  \end{aligned}\right.
\end{align*}
The tuning parameter $\gamma$ balances protection against bias with the pursuit of efficiency. Specifically, a smaller $\gamma$ shrinks $\delta_{t,N}$ to zero faster and extends protection to biases as small as $N^{\gamma/4-1/2}$, at the cost of returning to one more slowly, at rate $N^{-\gamma}$, when $\wh\tau_t$ is consistent. We use $\gamma=1/3$ in our numerical experiments, which guards against biases of order exceeding $N^{-5/12}$.
It is also of interest to consider \emph{local} model misspecification, in which the bias of $\wh\tau_t$ is of the same order $N^{-1/2}$ as its own standard error.
Further discussion and an alternative combination strategy under local misspecification are given in \Cref{sec_app:local}.

Let $\wh\lambda_{t,N}$ denote the empirical analog of $\lambda_{t,N}$. We define the resulting shrinkage estimator as $\wh\tau_t^{(s)}=\wh\tau_t(\wh\lambda_{t,N})$.
The following \Cref{thm:shrinkage_asy} establishes the asymptotic normality of $\wh\tau_t^{(s)}$ and the validity of inference under arbitrary misspecification.

\begin{Th}\label{thm:shrinkage_asy}
  Suppose that, for $t\in\{0,1\}$, (i) $\wt\tau_t$ is a $\sqrt{N}$-consistent estimator of $\tau_t$; (ii) $(\wt\tau_t,\wh\tau_t)$ are jointly asymptotically normal about their probability limits $(\tau_t,\tau_t^*)$, that is, $\sqrt N\{(\wt\tau_t-\tau_t),(\wh\tau_t-\tau_t^*)\}\trans\convd\calN(\0,\bSigma_t)$ with $\bSigma_t$ positive definite; and (iii) a consistent estimator $\wh\bSigma_t\convp\bSigma_t$ is available, so that $\wh V_{t,N}/V_{t,N}\convp1$.
  Write $\bc(\lambda)=(1-\lambda,\lambda)\trans$ and let $\lambda_t^{\dagger} = \lambda_t^{*}\,\mathbbm{1}(\tau_t^{*}=\tau_t)$ be the probability limit of $\wh\lambda_{t,N}$. Then, we have
    $\sqrt{N}\{\wh\tau_{t}^{(s)}-\tau_t\}\convd \calN(0,\,v_t)$,
    with $v_t = \bc(\lambda_t^{\dagger})\trans\bSigma_t\,\bc(\lambda_t^{\dagger})$.
  In addition, we have $\wh v_t=\bc(\wh\lambda_{t,N})\trans\wh\bSigma_t\,\bc(\wh\lambda_{t,N})\convp v_t$.
\end{Th}

\Cref{thm:shrinkage_asy} describes the adaptive behavior of $\wh\tau_t^{(s)}$. If $\wh\tau_t$ is inconsistent due to model misspecification, then $\bc(\lambda_t^{\dagger})=(1,0)$ and $\wh\tau_t^{(s)}$ degenerates to the trial-only estimator $\wt\tau_t$. If $\wh\tau_t$ is consistent for $\tau_t$, then $\bc(\lambda_t^{\dagger})=(1-\lambda_t^{*},\lambda_t^{*})$, and $\wh\tau_t^{(s)}$ aggregates $\wh\tau_t$ and $\wt\tau_t$ to incorporate information from the external control data to improve efficiency. Therefore, $\wh\tau_t^{(s)}$ adaptively interpolates between robustness and efficiency, and remains consistent regardless of whether the working models are correctly specified.
In practice, one may use the simple gold-standard estimators, $\wt\tau_1^\gs$ and $\wt\tau_0^\gs$, as baselines, which require no working models. Alternatively, one may use the trial-only doubly robust estimators obtained by solving estimating equations \eqref{eq:EIF1_rct}-\eqref{eq:EIF0_rct}, which additionally involve working models $\mu_t(\x;\bg_t^*)$. It is well known that consistency of the doubly robust estimator does not rely on correct specification of $\mu_t(\x;\bg_t^*)$, although efficiency does. The following corollary summarizes the shrinkage estimator's guaranteed consistency and its potential efficiency gains.

\begin{Cor}\label{cor:dominance}
  Suppose assumptions in \Cref{thm:consistency} and \Cref{thm:shrinkage_asy} hold. Then the shrinkage estimator $\wh\tau_{t}^{(s)}$ is consistent and asymptotically normal, with asymptotic variance no larger than that of the trial-based estimator $\wt\tau_t$, for any choice of working models $k(\x; \ba^*)$, $\rho(y_0,\x; \bb^*)$ and $\mu_t(\x;\bg_t^*)$.
\end{Cor}

\begin{Rem}
  We construct shrinkage estimators for the treatment and control arms, respectively. The motivation is that consistency of $\wh\tau_1$ only requires correct specification of $k(\x)$ and is therefore more plausible than consistency of $\wh\tau_0$, which requires correct specification of both $k(\x)$ and $\rho(\x,y_0)$. In case that $k(\x)$ is correctly specified but $\rho(\x,y_0)$ is not, the separate strategy can still leverage $\wh\tau_1$ to borrow information from the external data, while shrinking the control mean toward the trial-only estimator $\wt\tau_0$ to preserve consistency.
  It is also straightforward to adapt the construction when the estimand of interest is the treatment effect $\tau$. In this case, one can form a single shrinkage estimator that combines $\wt\tau$ and $\wh\tau$, and define an adaptive shrinkage weight via an analogous procedure. The resulting estimator remains consistent and is guaranteed to be at least as efficient as the baseline estimator $\wt\tau$.
\end{Rem}

\section{Numerical Studies}\label{sec:simulation}
In this section, we conduct three synthetic experiments and a real-data analysis to assess the finite-sample performance of the proposed estimators and compare them with the main existing approaches to external-control borrowing. The first simulation illustrates how the methods behave under increasing departures from exchangeability, the second quantifies the efficiency gains attainable from different types of information about the distributional shift, and the third assesses the robustness of the shrinkage estimator to model misspecification. Supplemental results are collected in \Cref{sec_app:numerical}.

For the trial, we generate $X\sim \calN(0,1)$ and $Y_t\mid x\sim \calN\{\mu_t(x),\sigma^2(x)\}$ with $\mu_0(x)=1+x+0.5x^2$, $\mu_1(x)=2+1.8x+0.5x^2$, and $\sigma^2(x)=0.5+0.5x^2$. The treatment assignment probability is $\pi=0.5$.
For the external controls, we generate $X\sim \calN(\mu_E,\sigma_E^2)$ with $(\mu_E,\sigma_E)=(0.5,1.2)$, which controls the magnitude of the covariate shift, and we let the distribution of the untreated outcome be the exponential tilt of its trial counterpart, $q(y_0\mid x)\propto \exp(\theta y_0)\,p(y_0\mid x)$, so that $Y\mid x\sim \calN\{\mu_0(x)+\theta\sigma^2(x),\,\sigma^2(x)\}$. The tilting parameter $\theta$ controls the magnitude of the concept shift, with $\theta=0.5$ as the default and $\theta=0$ corresponding to the scenario in which exchangeability holds.
Under this data-generating mechanism, the shift functions take the logistic forms $k(x;\ba)=\mathrm{expit}(\alpha_0+\alpha_1x+\alpha_2x^2)$ and $\rho(x,y_0;\bb)=\mathrm{expit}(\beta_0+\beta_1x+\beta_2x^2+\beta_3y_0)$ exactly, with $\beta_3=-\theta$. Throughout, we use $2000$ Monte Carlo replicates unless stated otherwise.

We compare the trial-only estimator $\wt\tau_t$ constructed from \eqref{eq:EIF1_rct}-\eqref{eq:EIF0_rct}, the EC-augmented estimator $\wh\tau_t$ and the shrinkage estimator $\wh\tau_t^{(s)}$ with four alternatives: the \emph{naive pooling} estimator, which treats the external controls as trial controls without adjustment; the \emph{exchangeability-efficient} estimator \citep{LiMiaoLuZhou2023}, which is semiparametrically efficient under conditional mean exchangeability; the \emph{adaptive selective borrowing} approach \citep{GaoYangShanYEEtAl2024}, which uses an adaptive lasso to borrow only the subset of external controls that are compatible with exchangeability; and the \emph{randomization-aware} approach \citep{KarlssonWangBartolomeisKrijtheEtAl2025}, which uses the external controls solely to construct the outcome regression while relying on the known trial randomization probability for identification.

\begin{figure}
\centering
\includegraphics[page=1,width=\textwidth]{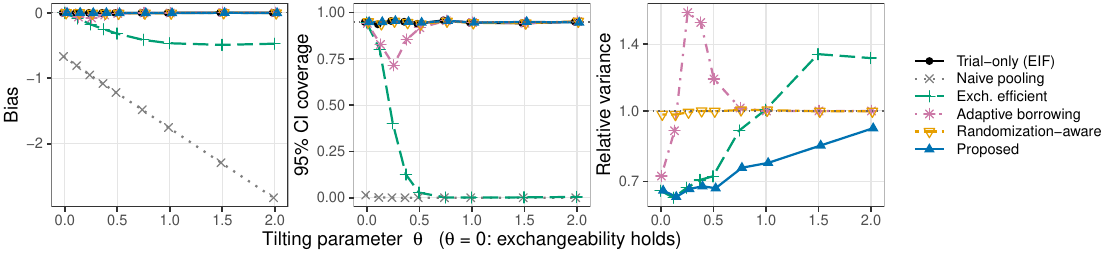}
\caption{Estimation of $\tau$ as the tilting parameter $\theta$ grows, with $n=500$ and $N=1500$. The proposed estimator treats $\theta$ as known. Relative variance is on a log scale against the trial-only estimator, with smaller values indicating greater efficiency.}
\label{fig:compare_concept}
\end{figure}

\textbf{Experiment I.}~
In this experiment, we vary $\theta$ to control the degree of violation of the exchangeability and treat its value as known a priori as in \citet{KimYu2011}. The proposed method can explicitly incorporate this information to adjust the external outcomes, whereas the competing methods do not.  \Cref{fig:compare_concept} reports the bias, coverage, and relative variance of $\wh\tau$ for a trial of size $n=500$ within a total sample of $N=1500$. Tables and additional results are deferred to \Cref{sec_app:numerical}.

Naive pooling is biased throughout. When exchangeability holds, i.e. $\theta=0$, all the remaining methods are unbiased with satisfactory coverage, and the methods of \citet{LiMiaoLuZhou2023}, \citet{GaoYangShanYEEtAl2024}, and our method exploiting the knowledge of $\theta=0$ all exhibit clear efficiency gains. As the departure from exchangeability grows, the exchangeability-efficient estimator becomes increasingly biased, as expected, and its coverage deteriorates. Adaptive selective borrowing shrinks the retained external sample to nearly an empty set, from $97.5\%$ at $\theta=0$ to $0.1\%$ once $\theta\ge0.75$; it therefore remains unbiased but achieves no efficiency gain, while exhibiting some numerical instability for intermediate values of $\theta$. The randomization-aware estimator is unbiased across all settings, but shows no appreciable efficiency gain.
In contrast, the proposed estimator uses $\theta$ directly through the working model $\rho(x,y_0;\bb)$, without any model for the conditional distribution of $Y_0$ beyond its mean, and remains unbiased with nominal coverage and a variance between $0.65$ and $0.92$ times that of the trial-only estimator across the whole range of $\theta$.

\textbf{Experiment II.}~
This experiment investigates numerically what information the proposed estimator is able to exploit. We consider three factors, namely the availability of an external validation sample, knowledge of the shape of the shift, and the size of the external sample relative to that of the trial.

\begin{figure}
\centering
\includegraphics[page=2,width=\textwidth]{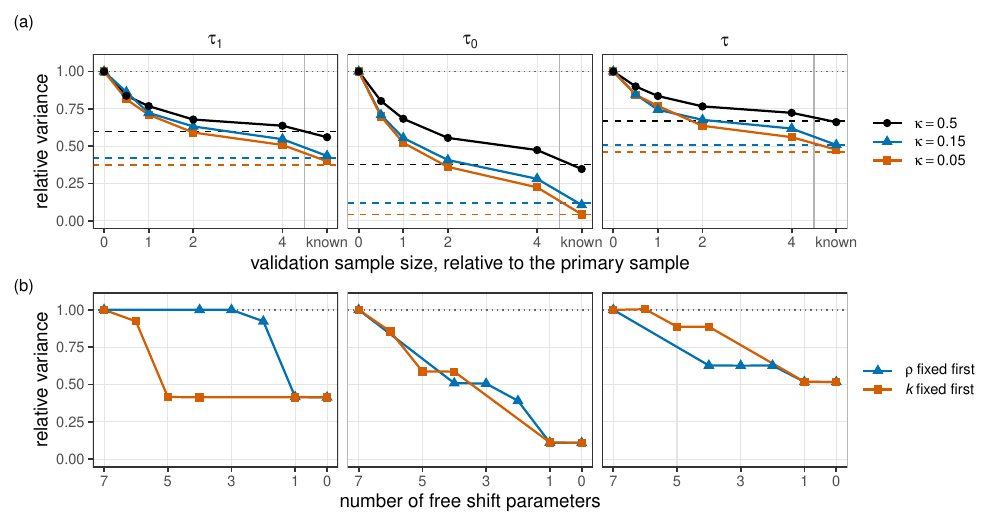}
\caption{Relative variance of the proposed augmented estimator as information about the shift increases, through (a) a validation sample of increasing size and (b) components of $k$ and $\rho$ fixed at their true values. Dashed lines in (a) indicate the known-shift limits given in \Cref{prop:var}.}
\label{fig:information}
\end{figure}

\Cref{fig:information}(a) displays the variance of the proposed augmented estimator relative to that of the trial-only estimator, plotted against the size of the validation sample for three values of the trial fraction $\kappa=n/N$. The efficiency gain increases steadily with the validation sample size and approaches the known-shift benchmark indicated by the dashed lines. A larger external sample likewise yields a greater gain. \Cref{fig:information}(b) displays the same quantity as individual components of $k$ and $\rho$ are successively treated as known, proceeding from all seven unknown to none along two distinct orderings. Additional knowledge of the shift model yields a further gain at every step, and the two orderings converge to the same benchmark.

\textbf{Experiment III.}~
In this experiment, we examine the robustness of the different estimators to potential model misspecification. We consider four scenarios, where S1 specifies $\mu_t$, $k$ and $\rho$ correctly, S2 misspecifies $\mu_t$ alone, S3 misspecifies $k$ and $\rho$ alone, and S4 misspecifies all three. Each misspecified shift model omits a quadratic term present in the true model, and the misspecified outcome model uses $|x|$ in place of $x$ as covariates. All estimators are provided with a validation sample of the same size as the primary sample.

\begin{figure}
\centering
\includegraphics[page=3,width=\textwidth]{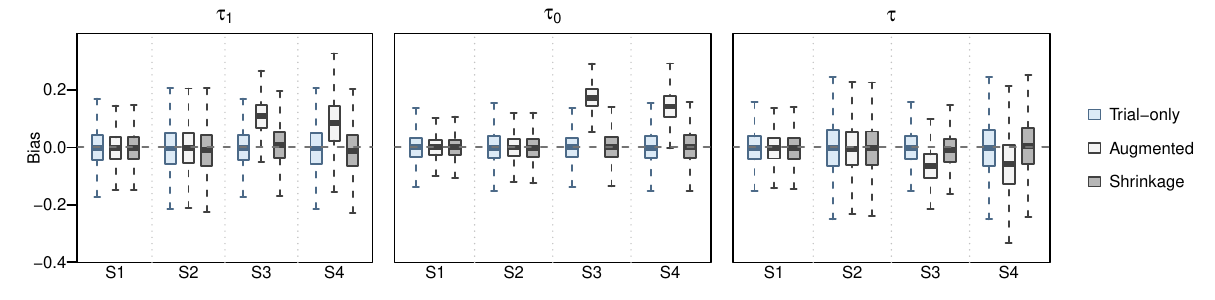}
\caption{Empirical bias under four working-model misspecification scenarios.}
\label{fig:robustness}
\end{figure}

\Cref{fig:robustness} presents boxplots of the trial-only estimator $\wt\tau_t$, the augmented estimator $\wh\tau_t$, and the shrinkage estimator $\wh\tau_t^{(s)}$ across the four scenarios. Under S1 and S2, all three estimators are centred at the true value, illustrating the robustness of $\wh\tau_t$ to misspecification of $\mu_t$, as established in \Cref{thm:consistency}. Under S3 and S4, in which $k$ and $\rho$ are misspecified, $\wh\tau_t$ is visibly biased, whereas $\wh\tau_t^{(s)}$ remains centred at the truth. These results indicate that the shrinkage estimator preserves consistency even when the models for the external data are misspecified.

\textbf{Real-data application.}~
We evaluate our method on the LaLonde training-program benchmark, combining data from the National Supported Work (NSW), a randomized job-training program, with an external control group from the Panel Study of Income Dynamics (PSID) \citep{Lalonde1986,DehejiaWahba1999}.  Detailed results are presented in \Cref{sec_app:realdata}.

\section{Conclusions}

\paragraph{Related work.}
External controls have long been used to improve trial precision, dating back to historical-control combination designs \citep{Pocock1976}. Under exchangeability between trial and external controls, a range of methods have been proposed to borrow EC information for trial inference \citep{DahabrehRobertsonTchetgenStuartEtAl2019,LiMiaoLuZhou2023,ValanciusPangZhuColeEtAl2024,wang2024evaluating}. Related historical-borrowing approaches assess compatibility or discount external information before borrowing, including test-then-pool methods \citep{VieleBerryNeuenschwanderAmzalEtAl2014} and Bayesian dynamic borrowing \citep{HobbsSargentCarlin2012,SchmidliGsteigerRoychoudhuryOHaganEtAl2014}.

When exchangeability is questionable, recent work has developed more guarded strategies for incorporating external controls. These include selective borrowing from compatible controls \citep{GaoYangShanYEEtAl2024,YangLiWu2025}, weighted or shrinkage-based combinations \citep{OberstDAmourChenWangEtAl2022,RosenmanBasseOwenBaiocchi2023,YangGaoZengWang2023}, and prognostic covariate adjustment that avoids direct outcome pooling \citep{SchulerWalshHallWalshEtAl2022,LiaoHojbjerre-FrandsenHubbardSchuler2025}. Randomization-aware combinations further preserve trial-based consistency and can protect against efficiency loss when EC information is unreliable \citep{wang2025robust,KarlssonWangBartolomeisKrijtheEtAl2025,bartolomeis2026efficient}.
Our work differs from these lines by explicitly exploiting the distribution shifts that void exchangeability, using them to calibrate EC data toward the trial population, and combining the resulting augmented estimator with a trial-only benchmark via adaptive shrinkage to retain trial-based guarantees.

More broadly, EC borrowing is connected to the broader literature on data fusion and integration, including efficient fusion \citep{LiLuedtke2023}, multi-source integration \citep{YangDing2020}, bias correction and double machine learning \citep{ParikhMorucciOrlandiRoyEtAl2025}, long-term effects \citep{GhassamiYangRichardsonShpitserEtAl2022}, and transfer learning \citep{WuYang2023}; see recent reviews for broader context \citep{ColnetMayerChenDiengEtAl2024,LinTarpEvans2024}.

\textbf{Position of our work in the literature.}~
Our work departs from existing approaches in three main ways.
First, while most prior methods rely on some form of exchangeability between RCT and the EC population, we treat distribution shift as a structural feature of the data rather than a nuisance to be assumed away, and we directly incorporate it into estimation through calibration equations.
Second, in contrast to guarded approaches that defensively attenuate EC influence via selection, downweighting, or shrinkage, our framework aims to actively exploit EC data under a transparent shift model; in particular, the adaptive shrinkage proposal provides a principled safeguard that recovers the trial-only benchmark under shift-model misspecification.
Third, whereas existing methods typically borrow EC information only for the control arm, our calibration-based augmentation leverages EC covariate information for both arms, so that, when properly aligned with the trial population, EC data can sharpen inference for treatment-arm functionals as well.

Methodologically, our framework connects to the broader literature on distribution shift and on data fusion and transportability in causal inference, specialized to the semiparametric efficiency calculus of trial inference.
The combination of explicit shift modeling, dual-arm borrowing, and provable efficiency dominance distinguishes our paper from prior work in the literature.

\appendix
\crefalias{section}{appendix}
\crefalias{subsection}{appendix}
\crefname{appendix}{Appendix}{Appendices}
\Crefname{appendix}{Appendix}{Appendices}
\renewcommand{\thefigure}{A.\arabic{figure}}
\renewcommand{\thetable}{A.\arabic{table}}
\setcounter{figure}{0}
\setcounter{table}{0}

\section{Local Misspecification and Alternative Bias-Aware Estimator}\label{sec_app:local}

In the conventional data-generating process, we pre-determine either $\tau_t^*=\tau_t$ or $\tau_t^*\neq\tau_t$, with the distinction fixed as the sample size $N$ grows. \emph{Local misspecification} is the case in between, with $\tau_t^*-\tau_t=hN^{-1/2}$ for a fixed $h$, so that the discrepancy has the same order as the standard error of $\wt\tau_t$ and a larger sample never settles which of the two holds. It becomes relevant in practice when the trial is small enough that $\tau_t^*$ and $\tau_t$ differ by little relative to the sampling error. The studentized discrepancy $T_{t,N}:=(\wh\tau_t-\wt\tau_t)/\sqrt{V_{t,N}}$ is then $O_p(1)$, so that $\delta_{t,N}\convp1$, $\wh\lambda_{t,N}\convp\lambda_t^*$, and
\begin{align*}
\sqrt N\{\wh\tau_t^{(s)}-\tau_t\}\convd\calN(\lambda_t^*h, v_t).
\end{align*}
Therefore, the coverage falls below the nominal level whenever $h\neq0$, though the point estimate remains $\sqrt N$-consistent.

One remedy is to bound the bias manually by the analyst. Suppose one is willing to assert $|\tau_t^*-\tau_t|\le B$. Let
\begin{align*}
\lambda_B=\lambda_t^*\,\frac{V_{t,N}}{V_{t,N}+B^2},
\end{align*}
which has the form of $\lambda_{t,N}=\delta_{t,N}\lambda_t^*$ with the data-driven factor $N^{-\gamma}T_{t,N}^4$ in
\begin{align*}
\delta_{t,N}=\left\{1+N^{-\gamma}T_{t,N}^4\right\}^{-1},
\end{align*}
replaced by the deterministic $B^2/V_{t,N}$. Let $\wh\sigma_t^2(\lambda)=\bc(\lambda)\trans\wh\bSigma_t\bc(\lambda)/N$ for the estimated variance of $\wh\tau_t(\lambda)$. Then, a bias-aware estimator is given by
\begin{align*}
\wh\tau_t^{(B)}=\wt\tau_t+\wh\lambda_B(\wh\tau_t-\wt\tau_t),~~\text{with}~~
\wh\lambda_B=\wh\lambda_t^*\,\frac{\wh V_{t,N}}{\wh V_{t,N}+B^2}.
\end{align*}
Based on this estimator, a confidence interval at nominal level $1-\alpha$ that explicitly accounts for the bias is
\begin{align*}
\wh\tau_t^{(B)}\pm \mathrm{cv}_{1-\alpha}\!\left\{\frac{\wh\lambda_B B}{\wh\sigma_t(\wh\lambda_B)}\right\}\wh\sigma_t(\wh\lambda_B),
\end{align*}
where $\mathrm{cv}_{1-\alpha}(r)$ denotes the $1-\alpha$ quantile of $|\calN(r,1)|$, the fixed-length critical value \citep{Donoho1994,ArmstrongKolesar2018}. By construction, it attains coverage at least $1-\alpha$ uniformly over $|\tau_t^*-\tau_t|\le B$.

The bias-aware estimator is motivated differently from the shrinkage estimator proposed in \Cref{sec:shrinkage}. Specifically, the shrinkage estimator decides whether, and by how much, to borrow from the external data based on the observed discrepancy between the two estimators. This strategy cannot detect a local bias by construction, as it is indistinguishable from sampling noise at the relevant scale. In contrast, the bias-aware estimator $\lambda_B$ is built on how much bias one is prepared to tolerate. It yields valid inference throughout the range at the cost of a weight fixed by assumption rather than learned, and of no protection beyond the bound. Since $B$ traces the entire path from the oracle combination $\wh\tau_t(\lambda_t^*)$ at $B=0$ to the trial-only estimator as $B\to\infty$, it is recommended to report the estimate and the interval across a range of $B$ for sensitivity analysis.

\section{Additional Numerical Results}\label{sec_app:numerical}
This appendix collects the full numerical summaries underlying the figures of \Cref{sec:simulation}. Throughout, ``Rel.\ var.'' denotes the ratio of the empirical variance of a given estimator to that of the trial-only estimator, which is the quantity the figures plot, ``Rel.\ MSE'' the corresponding ratio of mean squared errors, $\overline{\mathrm{SE}}$ the root mean square of the estimated standard errors. Monte Carlo standard errors for the coverage entries are at most $0.005$ with $2000$ replicates.

\begin{figure}[htbp]
\centering
\includegraphics[page=4,width=\textwidth]{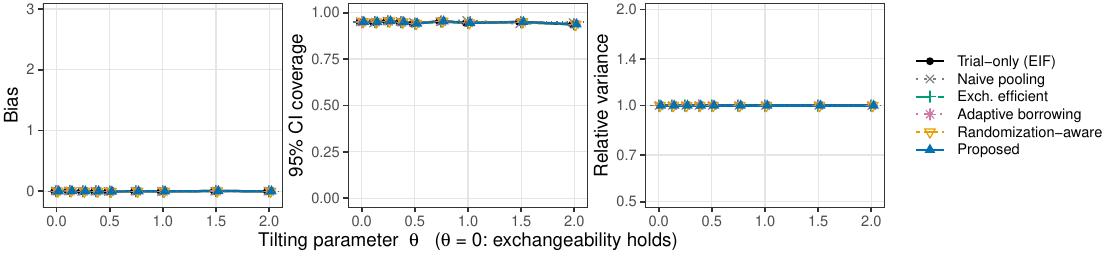}\\[0.5ex]
\includegraphics[page=5,width=\textwidth]{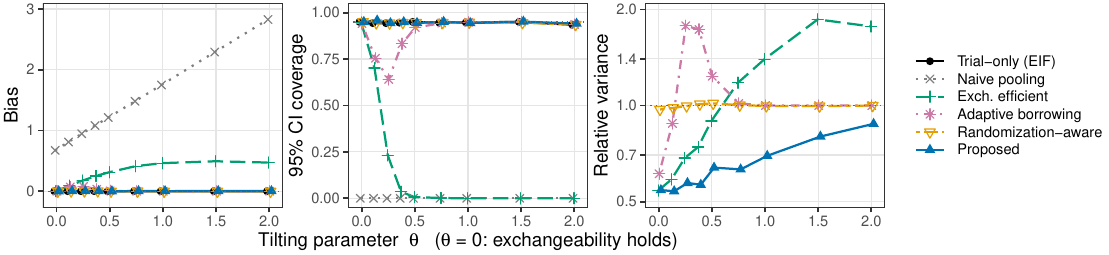}
\caption{Experiment I. Arm-specific estimands for $\tau_1$ (top) and $\tau_0$ (bottom).}
\label{fig:app_E2_arms}
\end{figure}

In summary, \Cref{tab:app_E2} and \Cref{fig:app_E2_arms} accompany the first experiment, reporting the values underlying \Cref{fig:compare_concept} together with results for the two arm-specific estimands. \Cref{tab:app_E1,tab:app_E3,tab:app_E4b} accompany the second experiment, where \Cref{tab:app_E1} records the bias and coverage, and the remaining two report the values underlying its two panels. \Cref{tab:app_E5} accompanies the third experiment.

\begin{table}[htbp]
\centering
\caption{Experiment I. Comparison for $\tau$ as the tilting parameter $\theta$ grows ($n=500$, $N=1500$), corresponding to \Cref{fig:compare_concept}. The proposed estimator treats $\theta$ as known.}
\label{tab:app_E2}
\footnotesize
\setlength{\tabcolsep}{4pt}
\begin{tabular}{llrrrrrrrrr}
\toprule
Method & Metric & 0.00 & 0.12 & 0.25 & 0.38 & 0.50 & 0.75 & 1.00 & 1.50 & 2.00 \\
\midrule
\multirow{3}{*}{Trial-only (EIF)} & Bias & 0.001 & 0.003 & -0.001 & -0.001 & -0.002 & -0.001 & -0.001 & 0.004 & -0.001 \\
 & Coverage & 0.949 & 0.936 & 0.955 & 0.951 & 0.940 & 0.958 & 0.946 & 0.943 & 0.951 \\
 & Rel.\ var. & 1.00 & 1.00 & 1.00 & 1.00 & 1.00 & 1.00 & 1.00 & 1.00 & 1.00 \\
\multirow{3}{*}{Naive pooling} & Bias & -0.668 & -0.808 & -0.949 & -1.081 & -1.218 & -1.482 & -1.753 & -2.292 & -2.827 \\
 & Coverage & 0.014 & 0.002 & 0.001 & 0.000 & 0.000 & 0.000 & 0.000 & 0.000 & 0.000 \\
 & Rel.\ var. & 2.64 & 2.52 & 2.42 & 2.52 & 2.52 & 2.87 & 2.89 & 3.01 & 3.59 \\
\multirow{3}{*}{Exch. efficient} & Bias & 0.000 & -0.087 & -0.176 & -0.251 & -0.317 & -0.414 & -0.465 & -0.490 & -0.472 \\
 & Coverage & 0.947 & 0.802 & 0.402 & 0.126 & 0.028 & 0.002 & 0.001 & 0.002 & 0.004 \\
 & Rel.\ var. & 0.67 & 0.65 & 0.68 & 0.71 & 0.72 & 0.91 & 1.01 & 1.33 & 1.31 \\
\multirow{3}{*}{Adaptive borrowing} & Bias & 0.000 & -0.065 & -0.079 & -0.033 & -0.012 & -0.001 & -0.001 & 0.004 & -0.001 \\
 & Coverage & 0.942 & 0.828 & 0.716 & 0.855 & 0.917 & 0.956 & 0.947 & 0.942 & 0.952 \\
 & Rel.\ var. & 0.72 & 0.90 & 1.64 & 1.56 & 1.17 & 1.01 & 1.00 & 1.00 & 1.00 \\
\multirow{3}{*}{Randomization-aware} & Bias & 0.000 & -0.000 & -0.007 & -0.007 & -0.009 & -0.006 & -0.006 & 0.001 & -0.003 \\
 & Coverage & 0.950 & 0.939 & 0.953 & 0.946 & 0.938 & 0.953 & 0.945 & 0.943 & 0.953 \\
 & Rel.\ var. & 0.98 & 0.98 & 0.99 & 1.00 & 1.00 & 1.00 & 1.00 & 1.00 & 1.00 \\
\multirow{3}{*}{Proposed} & Bias & 0.000 & 0.002 & -0.003 & -0.001 & -0.004 & -0.003 & -0.004 & -0.000 & -0.003 \\
 & Coverage & 0.949 & 0.949 & 0.955 & 0.950 & 0.946 & 0.956 & 0.946 & 0.950 & 0.947 \\
 & Rel.\ var. & 0.67 & 0.65 & 0.67 & 0.68 & 0.68 & 0.75 & 0.77 & 0.84 & 0.92 \\
\bottomrule
\end{tabular}
\end{table}

\begin{table}[htbp]
\centering
\caption{Experiment II. Bias, dispersion and coverage as the amount of information about the shift varies at $n=1500$ and $N=3500$. Here $\xi$ is the size of the validation sample relative to the primary sample. \Cref{fig:information} plots relative variance only.}
\label{tab:app_E1}
\footnotesize
\setlength{\tabcolsep}{4pt}
\begin{tabular}{lcrrrrrr}
\toprule
Method & $\xi$ & Bias & SD & $\overline{\mathrm{SE}}$ & SE/SD & Coverage & Rel.\ var. \\
\midrule
\multicolumn{8}{l}{\emph{$\tau_1$}} \\
Trial-only (EIF) & -- & -0.0021 & 0.064 & 0.062 & 0.96 & 0.939 & 1.000 \\
Proposed & 0 & -0.0021 & 0.064 & 0.062 & 0.96 & 0.939 & 1.000 \\
 & 0.5 & -0.0019 & 0.059 & 0.057 & 0.97 & 0.950 & 0.839 \\
 & 1 & -0.0028 & 0.057 & 0.055 & 0.97 & 0.934 & 0.777 \\
 & 2 & -0.0006 & 0.053 & 0.052 & 0.98 & 0.949 & 0.678 \\
 & 4 & -0.0014 & 0.051 & 0.050 & 0.97 & 0.944 & 0.635 \\
Proposed (known shift) & -- & -0.0009 & 0.047 & 0.046 & 0.98 & 0.945 & 0.541 \\
\midrule
\multicolumn{8}{l}{\emph{$\tau_0$}} \\
Trial-only (EIF) & -- & -0.0007 & 0.050 & 0.048 & 0.96 & 0.939 & 1.000 \\
Proposed & 0 & -0.0007 & 0.050 & 0.048 & 0.96 & 0.939 & 1.000 \\
 & 0.5 & -0.0003 & 0.044 & 0.043 & 0.98 & 0.942 & 0.752 \\
 & 1 & -0.0011 & 0.040 & 0.039 & 0.98 & 0.948 & 0.631 \\
 & 2 & 0.0002 & 0.035 & 0.036 & 1.01 & 0.956 & 0.494 \\
 & 4 & 0.0001 & 0.033 & 0.033 & 1.00 & 0.951 & 0.424 \\
Proposed (known shift) & -- & 0.0005 & 0.026 & 0.027 & 1.04 & 0.959 & 0.274 \\
\midrule
\multicolumn{8}{l}{\emph{$\tau$}} \\
Trial-only (EIF) & -- & -0.0014 & 0.057 & 0.056 & 0.97 & 0.949 & 1.000 \\
Proposed & 0 & -0.0014 & 0.057 & 0.056 & 0.97 & 0.949 & 1.000 \\
 & 0.5 & -0.0016 & 0.053 & 0.052 & 0.98 & 0.946 & 0.858 \\
 & 1 & -0.0017 & 0.051 & 0.050 & 0.98 & 0.942 & 0.805 \\
 & 2 & -0.0007 & 0.050 & 0.048 & 0.97 & 0.946 & 0.754 \\
 & 4 & -0.0015 & 0.048 & 0.047 & 0.97 & 0.951 & 0.711 \\
Proposed (known shift) & -- & -0.0014 & 0.046 & 0.044 & 0.97 & 0.941 & 0.640 \\
\bottomrule
\end{tabular}
\end{table}

\begin{table}[htbp]
\centering
\caption{Experiment II. Variance relative to trial-only as the validation size $\xi$ and the trial fraction $\kappa$ vary, with the \Cref{prop:var} limit in the last column.}
\label{tab:app_E3}
\footnotesize
\setlength{\tabcolsep}{4pt}
\begin{tabular}{crrrrrrrr}
\toprule
Estimand & $\kappa$ & $\xi=0$ & $\xi=0.5$ & $\xi=1$ & $\xi=2$ & $\xi=4$ & known & Prop.~1 \\
\midrule
$\tau_1$ & 0.500 & 1.000 & 0.838 & 0.768 & 0.678 & 0.637 & 0.560 & 0.599 \\
 & 0.150 & 1.000 & 0.862 & 0.723 & 0.632 & 0.547 & 0.435 & 0.423 \\
 & 0.050 & 1.000 & 0.814 & 0.711 & 0.592 & 0.508 & 0.399 & 0.373 \\
\midrule
$\tau_0$ & 0.500 & 1.000 & 0.802 & 0.683 & 0.555 & 0.474 & 0.347 & 0.377 \\
 & 0.150 & 1.000 & 0.709 & 0.555 & 0.407 & 0.282 & 0.107 & 0.120 \\
 & 0.050 & 1.000 & 0.699 & 0.522 & 0.361 & 0.225 & 0.044 & 0.043 \\
\midrule
$\tau$ & 0.500 & 1.000 & 0.900 & 0.836 & 0.766 & 0.723 & 0.661 & 0.667 \\
 & 0.150 & 1.000 & 0.847 & 0.744 & 0.676 & 0.618 & 0.509 & 0.510 \\
 & 0.050 & 1.000 & 0.844 & 0.769 & 0.636 & 0.560 & 0.477 & 0.459 \\
\bottomrule
\end{tabular}
\end{table}

\begin{table}[htbp]
\centering
\caption{Experiment II. Variance relative to trial-only when part of the shift is fixed at its true value, corresponding to \Cref{fig:information}(b).}
\label{tab:app_E4b}
\footnotesize
\setlength{\tabcolsep}{4pt}
\begin{tabular}{crrr rrr rrr}
\toprule
Free parameters & \multicolumn{3}{c}{$\tau_1$} & \multicolumn{3}{c}{$\tau_0$} & \multicolumn{3}{c}{$\tau$} \\
\cmidrule(lr){2-4}\cmidrule(lr){5-7}\cmidrule(lr){8-10}
in $(k,\rho)$ & 0.50 & 0.15 & 0.05 & 0.50 & 0.15 & 0.05 & 0.50 & 0.15 & 0.05 \\
\midrule
$(3,4)$ & 1.000 & 1.000 & 1.000 & 1.000 & 1.000 & 1.000 & 1.000 & 1.000 & 1.000 \\
$(2,4)$ & 0.931 & 0.923 & 0.919 & 0.916 & 0.854 & 0.884 & 1.001 & 1.004 & 1.024 \\
$(1,4)$ & 0.584 & 0.417 & 0.394 & 0.710 & 0.588 & 0.603 & 0.926 & 0.887 & 0.921 \\
$(3,1)$ & 1.000 & 1.000 & 1.000 & 0.683 & 0.509 & 0.466 & 0.743 & 0.627 & 0.576 \\
$(0,4)$ & 0.560 & 0.414 & 0.394 & 0.694 & 0.586 & 0.603 & 0.922 & 0.887 & 0.921 \\
$(3,0)$ & 1.000 & 1.000 & 1.000 & 0.654 & 0.506 & 0.465 & 0.729 & 0.626 & 0.580 \\
$(2,0)$ & 0.931 & 0.923 & 0.919 & 0.590 & 0.391 & 0.336 & 0.729 & 0.627 & 0.586 \\
$(1,0)$ & 0.584 & 0.417 & 0.394 & 0.363 & 0.111 & 0.046 & 0.665 & 0.518 & 0.476 \\
$(0,1)$ & 0.560 & 0.414 & 0.394 & 0.379 & 0.113 & 0.044 & 0.674 & 0.518 & 0.473 \\
$(0,0)$ & 0.560 & 0.414 & 0.394 & 0.347 & 0.110 & 0.046 & 0.661 & 0.517 & 0.476 \\
\bottomrule
\end{tabular}
\end{table}

\begin{table}[htbp]
\centering
\caption{Experiment III. Robustness under the four specification scenarios at $n=1500$, $N=3500$, corresponding to \Cref{fig:robustness}.}
\label{tab:app_E5}
\footnotesize
\setlength{\tabcolsep}{4pt}
\begin{tabular}{clrrr rrr}
\toprule
Scenario & Estimator & \multicolumn{3}{c}{Bias} & \multicolumn{3}{c}{Coverage} \\
\cmidrule(lr){3-5}\cmidrule(lr){6-8}
 & & $\tau_1$ & $\tau_0$ & $\tau$ & $\tau_1$ & $\tau_0$ & $\tau$ \\
\midrule
S1 & Trial-only & -0.0021 & -0.0007 & -0.0014 & 0.939 & 0.939 & 0.949 \\
 & Augmented & -0.0023 & -0.0004 & -0.0018 & 0.943 & 0.952 & 0.950 \\
 & Shrinkage & -0.0028 & -0.0012 & -0.0016 & 0.937 & 0.941 & 0.945 \\
\midrule
S2 & Trial-only & -0.0027 & -0.0006 & -0.0021 & 0.949 & 0.946 & 0.949 \\
 & Augmented & -0.0030 & -0.0008 & -0.0021 & 0.953 & 0.951 & 0.939 \\
 & Shrinkage & -0.0099 & -0.0016 & -0.0017 & 0.944 & 0.941 & 0.938 \\
\midrule
S3 & Trial-only & -0.0021 & -0.0007 & -0.0014 & 0.939 & 0.939 & 0.949 \\
 & Augmented & 0.1078 & 0.1719 & -0.0642 & 0.558 & 0.032 & 0.785 \\
 & Shrinkage & 0.0077 & 0.0016 & -0.0100 & 0.914 & 0.939 & 0.938 \\
\midrule
S4 & Trial-only & -0.0027 & -0.0006 & -0.0021 & 0.949 & 0.946 & 0.949 \\
 & Augmented & 0.0845 & 0.1427 & -0.0582 & 0.839 & 0.253 & 0.918 \\
 & Shrinkage & -0.0104 & 0.0029 & 0.0063 & 0.934 & 0.940 & 0.943 \\
\bottomrule
\end{tabular}
\end{table}

\section{Real Data Application}\label{sec_app:realdata}

We evaluate our method on the LaLonde training-program benchmark, combining data from the National Supported Work (NSW), a randomized job-training program, with an external control group from the Panel Study of Income Dynamics (PSID) \citep{Lalonde1986,DehejiaWahba1999}. The outcome is the change in earnings between 1975 and 1978, defined as $\texttt{re\_dif}=\texttt{re78}-\texttt{re75}$. We adjust for six pre-treatment covariates: $\texttt{Age}$, $\texttt{Education}$, $\texttt{Black}$, $\texttt{Hispanic}$, $\texttt{Married}$, and $\texttt{Nodegree}$. To mimic a validation setting, we randomly split the data into two equal parts; one is used for primary estimation, and the other is treated as a validation sample.

\begin{figure}[htbp]
\centering
\includegraphics[page=6,width=.85\textwidth]{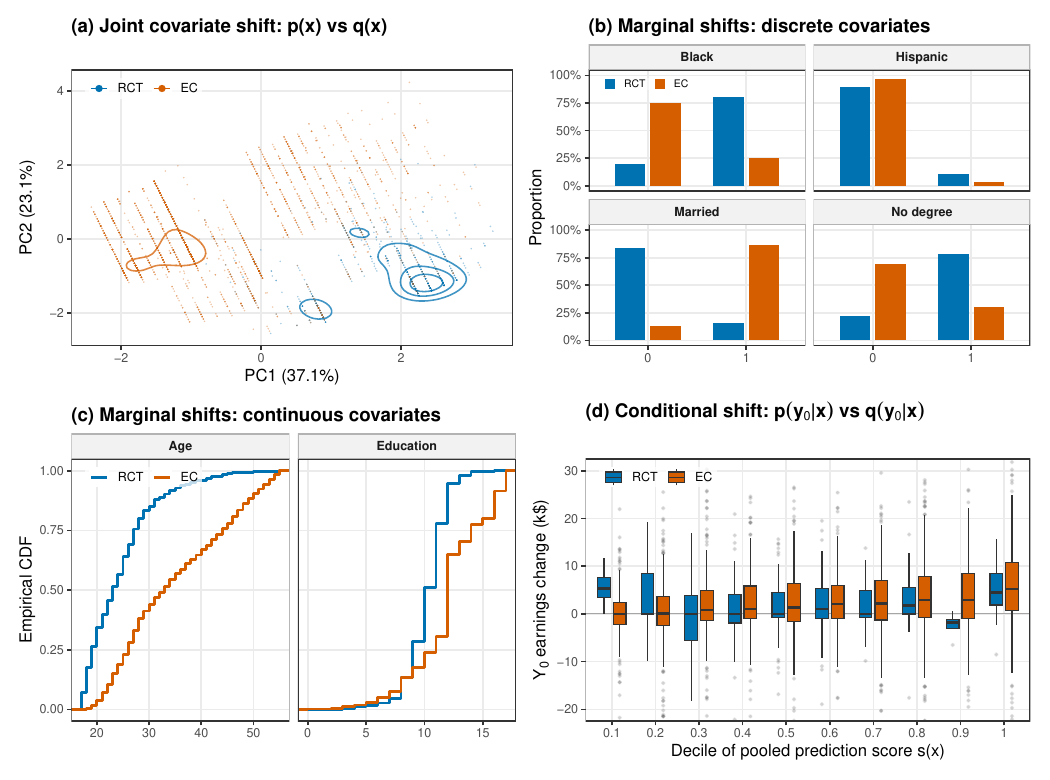}
\caption{
Visualization of distribution shifts between the NSW RCT and PSID EC samples.
(a) PCA projection of pooled covariates; percentages indicate variance explained, and contours show group-specific densities.
(b)-(c) Marginal shifts in discrete covariates via group-wise proportions and in continuous covariates via Empirical CDFs.
(d) Boxplots of observed $Y_0$ within deciles of the pooled prediction score $s(\x)=\wh\E(Y_0\mid\x)$.
}
\label{fig:real_distribution_shift}
\end{figure}

It is worth noting that the NSW trial and the PSID external controls are drawn from distinct data sources, with different sampling frames, collection periods, and measurement protocols, so that distribution shift between them is to be expected. \Cref{fig:real_distribution_shift} illustrates this shift. Panel~(a) shows the joint covariate distribution by projecting the mixed covariates onto their leading principal components, revealing separation between the two samples in covariate space. Panels~(b)-(c) further show marginal covariate shift. Panel~(d) examines conditional shift by grouping observations into deciles of a pooled covariate-based prediction score and comparing the outcome distributions within each decile. Differences in the boxplots within the same score stratum suggest that the conditional outcome distributions $p(y_0\mid \x)$ and $q(y_0\mid \x)$ differ across the two samples.

Exchangeability is therefore unlikely to hold in this application, which motivates applying the proposed procedure to model the discrepancy between the two data sources explicitly. In particular, we adopt logistic working models for $k(\x)$ and $\rho(\x,y_0)$, and estimate these nuisance components from the validation set. Table~\ref{tab:real_estimates} reports the resulting estimates, standard deviations, and 95\% confidence intervals for the trial-only estimator $\wt\tau_t$, the augmented estimator $\wh\tau_t$, and the shrinkage estimator $\wh\tau_t^{(s)}$. Without a validation set, directly incorporating EC data does not uniformly reduce variability relative to the trial-only estimator: the SDs of $\wh{\tau}_1$ and $\wh{\tau}$ are slightly larger than those of the corresponding trial-only baselines $\wt{\tau}_1$ and $\wt{\tau}$, whereas the SD of $\wh{\tau}_0$ is slightly smaller. By contrast, the shrinkage estimator retains the theoretical efficiency safeguard, with SDs no larger than those of the trial-only baselines across all three estimands. With a validation set, the EC-augmented estimators yield clearer variance reductions, especially for $\tau_0$ and $\tau$, and the shrinkage estimator preserves these gains while maintaining its efficiency guarantee.

\begin{table}[htbp]
\centering
\caption{
Real-data analysis results for the NSW trial and PSID external control samples.
}
\label{tab:real_estimates}
\renewcommand{\arraystretch}{1.15}
\resizebox{\ifdim\width>\linewidth\linewidth\else\width\fi}{!}{
\begin{tabular}{clccccc}
\toprule
&
& \multirow{2}{*}{Trial-only}
& \multicolumn{2}{c}{Without validation set}
& \multicolumn{2}{c}{With validation set} \\
\cmidrule(lr){4-5} \cmidrule(lr){6-7}
Estimand
&
&
& Augmented
& Shrinkage
& Augmented
& Shrinkage \\
\midrule
\multirow{2}{*}{$\tau_1$}
& Est. (SD)
& 2.418 {\small (0.551)}
& 2.399 {\small (0.568)}
& 2.444 {\small (0.543)}
& 2.416 {\small (0.546)}
& 2.411 {\small (0.542)} \\
& 95\% CI
& {\small [1.338, 3.498]}
& {\small [1.286, 3.512]}
& {\small [1.380, 3.508]}
& {\small [1.346, 3.486]}
& {\small [1.349, 3.473]} \\
\midrule
\multirow{2}{*}{$\tau_0$}
& Est. (SD)
& 1.567 {\small (0.436)}
& 1.686 {\small (0.429)}
& 1.787 {\small (0.427)}
& 1.647 {\small (0.355)}
& 1.653 {\small (0.355)} \\
& 95\% CI
& {\small [0.712, 2.422]}
& {\small [0.845, 2.527]}
& {\small [0.950, 2.624]}
& {\small [0.951, 2.343]}
& {\small [0.957, 2.349]} \\
\midrule
\multirow{2}{*}{$\tau$}
& Est. (SD)
& 0.851 {\small (0.702)}
& 0.713 {\small (0.711)}
& 0.907 {\small (0.701)}
& 0.769 {\small (0.649)}
& 0.759 {\small (0.648)} \\
& 95\% CI
& {\small [-0.524, 2.227]}
& {\small [-0.681, 2.107]}
& {\small [-0.467, 2.280]}
& {\small [-0.503, 2.041]}
& {\small [-0.512, 2.030]} \\
\bottomrule
\end{tabular}
}
\end{table}

\clearpage
\begingroup
\small
\setlength{\bibsep}{4pt}

\endgroup

\end{document}